\documentclass[11pt]{article}
\usepackage[textwidth=15.2cm,textheight=22cm]{geometry}
\usepackage{amsmath,amssymb}
\usepackage{latexsym}
\usepackage{multicol}
\usepackage{graphicx}
\usepackage{bm}
\usepackage{tikz}
\usetikzlibrary{shapes.geometric, arrows}
\usepackage{cite}
\usepackage{tikz}
\usetikzlibrary{trees}
\usepackage{varwidth}
\usepackage{pifont}
\usepackage{cancel}
\usepackage{centernot}
\usepackage{mathtools}
\usepackage[utf8]{inputenc}
\usepackage{longtable}
\usepackage{float}
\usepackage{simpler-wick}
\usepackage{xcolor}
\usepackage{dynkin-diagrams}
\newcommand{\xdownarrow}[1]{%
  {\left\downarrow\vbox to #1{}\right.\kern-\nulldelimiterspace}
}
\usetikzlibrary{shapes.geometric, arrows, shadows}

\tikzstyle{forces} = [rectangle, rounded corners, minimum width=2cm, minimum height=0.8cm,text centered, draw=black]
\tikzstyle{spin} = [rectangle, rounded corners, minimum width=1.3cm, minimum height=0.8cm,text centered, draw=black]
\tikzstyle{theory} = [rectangle, rounded corners, minimum width=2cm, minimum height=0.8cm,text centered, draw=gray]
\tikzstyle{arrow} = [thick,->,>=stealth]
\tikzstyle{arrow1} = [thick,->,>=stealth, color=red]
\tikzstyle{line} = [draw, -latex']

\allowdisplaybreaks[1]

\newcommand{\del}{\partial}
\newcommand{\be}{\begin{equation}}
\newcommand{\ee}{\end{equation}}
\newcommand{\ba}{\begin{eqnarray}}
\newcommand{\ea}{\end{eqnarray}}
\newcommand\fr[1]{\frac{1}{#1}}

\newcommand{\rom}[1]{\uppercase\expandafter{\romannumeral #1\relax}}
\newcommand{\nbar}[1]{\overline{#1}}

\def\ba{\bar A}

\def\beq{\begin{equation}}
\def\eeq{\end{equation}}

\newcommand{\nn}{\nonumber}

\newcommand{\ndt}{\noindent}

\def\bea{\begin{eqnarray}}
\def\eea{\end{eqnarray}}
\def\beas{\begin{eqnarray*}}
\def\eeas{\end{eqnarray*}}
\def\sla{\raise.15ex\hbox{$/$}\kern-.57em}

\def\parp{\partial^+}

\def\spa#1.#2{\left\langle#1\,#2\right\rangle}
\def\spb#1.#2{\left[#1\,#2\right]}

\date{}
\begin{document}
\begin{titlepage}
\begin{flushright}    
{\small $\,$}
\end{flushright}
\vskip 1cm
\vskip 1cm
\centerline{\Large{\bf{Exceptional symmetries in light-cone superspace}}}
\vskip 0.3 cm
\centerline{Sudarshan Ananth and Nipun Bhave}
\vskip 0.3cm
\centerline{\it {Indian Institute of Science Education and Research}}
\centerline{\it {Pune 411008, India}}
\vskip 1.5cm
\centerline{\bf {Abstract}}
\vskip .4cm
\ndt We construct maximal supergravity in five-dimensions by `oxidizing' the four-dimensional $\mathcal{N}=8$ theory. The relevant symmetries, the unitary symplectic group $USp(8)$ and the exceptional group $E_6$, are both presented in light-cone superspace and their connections with $SU(8)$ and $E_7$ highlighted. We explain a procedure to derive higher-point interaction vertices in both the 4- and 5-dimensional supergravity theories using exclusively the exceptional symmetries. Specific forms for the quartic and quintic interaction vertices in light-cone superspace are derived.

\vskip .5cm
\ndt 
\vfill
\end{titlepage}

\section{Introduction}

\ndt A striking feature of maximally supersymmetric theories of gravity in three, four and five-dimensions is the appearance of global exceptional symmetries. Minimal supersymmetry in higher dimensions always induces extended supersymmetry in a dimensionally-reduced theory and these theories, all arise from the same progenitor, the eleven-dimensional supergravity theory.  While the $d=11$ parent theory partly explains the origins of large global symmetries in the reduced models, it does not do so completely, relegating them to the bucket of `hidden' symmetries~\cite{Ananth:2018hif,Samtleben:2023nwk}. Starting from maximal supergravity in $d=11$, the model in $(11-n)$ dimensions exhibits a global $E_{n(n)}$ symmetry~\cite{Cremmer:1978ds,Cremmer:1979up,Cremmer:1978km,Cremmer:1979uq,Cremmer:1980gs}. The associated Dynkin diagrams starting from $E_8$ on the left are shown below (for $n\le 5$, the exceptional groups degenerate into the classical ones)
\vskip 0.3cm
\begin{figure}[h]
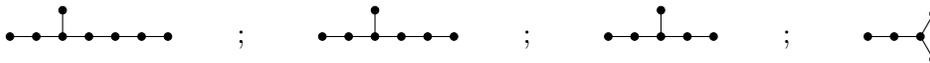

\dynkin E8   \;\;\;\;\; ; \;\;\;\;\;\;   \dynkin E7  \;\;\;\;\; ; \;\;\;\;\;\;  \dynkin E6  \;\;\;\;\; ; \;\;\;\;\;\; \dynkin D5
\caption{Dynkin diagrams of $E_8$,$E_7$, $E_{6}$ and $D_5$ }
\end{figure}

\ndt There has been considerable work on investigating the origins of these exceptional symmetries. Much of the literature suggests, strongly, that the eleven-dimensional parent has considerable information about the exceptional symmetries that emerge in its lower dimensional descendants~\cite{deWit:1986mz,deWit:1985iy,Damour:2002et,Damour:2002cu,West:2001as,Nicolai:1998gi,Ananth:2016abv,Ananth:2017nbi}. 

\vskip 0.3cm

\ndt A lot of the interest in these exceptional symmetries has stemmed from advances in scattering amplitude methods and with them, discoveries that $\mathcal N=8$ supergravity in $d=4$ is far better behaved in the ultra-violet than previously believed~\cite{Bern:2007hh,Bern:2008pv,Bern:2009kd,Bern:2011qn,Bern:2023zkg}. Whether the theory is ultra-violet finite to all orders remains an open question although the evidence suggests this is unlikely. The exact role of the exceptional symmetries in the improved finiteness properties remains unclear although progress has been made in understanding the possible counter terms~\cite{Howe:1980th,Howe:2002ui,Bossard:2011ij,Beisert:2010jx,Kallosh:2008rr,Kallosh:2011dp,Kallosh:2023css}. For an overview of progress in understanding maximal supergravity, see~\cite{Nicolai:2024hqh}.
\vskip 0.3cm
\ndt It is very possible that the spacetime group also has a large role to play in the finiteness saga~\cite{Bern:2007xj,Curtright:1982jz}. For example, Curtwright in \cite{Curtright:1982jz} conjectured that the ultraviolet finiteness of a supersymmetric theory in a given dimension could be attributed to the matching of Dynkin indices of the bosonic and fermionic representations of the massless little group in its parent theory. For the four-dimensional theory, this would suggest focussing on $SO(9)$, the little group in eleven-dimensions. The relevant Dynkin indices of $SO(9)$ are~\cite{Ramond:2001ud}
 \begin{longtable} [c] {| c | c | c | c |}
\hline
	Rep & (1001) & (2000) & (0010)\\
\hline
	$I_0$  & 128 &  44 & 84 \\
\hline
	$I_2$ & 256 & 88 & 168\\
\hline
	$I_4$ & 640 & 232 & 408\\
\hline
	$I_6$ & 1792 & 712 & 1080\\
\hline
	$I_8$ & 5248 & 2440 & 3000\\
\hline
\end{longtable}
 \vskip 0.3cm
 \ndt The mismatch of the eighth-order indices points to a possible divergence at three loops (at the earliest). Light-cone superspace offers a unified framework to study both symmetries and finiteness within one formalism. Originally developed for proving the ultra-violet finitess of $\mathcal N=4$ superYang-Mills theory~\cite{Brink:1982wv, Mandelstam:1982cb}, it has also proven useful to study the $\mathcal N=8$ theory~\cite{Ananth:2018mst}. 
\vskip 0.3cm
\ndt While there has been considerable work on four-dimensional supergravity, there are far fewer results pertaining to the five-dimensional theory~\cite{deWit:2004nw} particularly in this formalism~\cite{Ananth:2018mst}. This is an interesting theory in its own right because it is expected to be finite up to four loops~\cite{Bern:2009kd} and further, the global symmetries are symmetries of the entire action in odd dimensions~\cite{Cremmer:1980gs} as opposed to being just symmetries of the equations of motion - in theories in even dimensions\footnote{This is a more nuanced issue when working in the light-cone gauge since the formulation closely mimics an on-shell approach}. 
\vskip 0.3cm
\ndt We focus on two broad issues in this paper. First, we construct the five-dimensional supergravity theory to cubic order, starting from the four-dimensional $\mathcal N=8$ theory. We present the relevant symmetry algebras, $USp(8)$ and $E_{6(6)}$, in light-cone superspace. Second, we explain how these exceptional symmetries can play a crucial role in deriving interaction vertices in light-cone superspace. This is illustrated by obtaining forms for the quartic and quintic interaction vertices in both four and five dimensions. 
\vskip 0.3cm

\section{Supergravity: from $d=11$ to $d=5$}

\ndt The physical degrees of freedom of $\mathcal{N}=1$ supergravity in eleven-dimensions, which is also the low energy limit of M-theory~\cite{Hull:1994ys,Witten:1995ex}, are governed by the little group $SO(9)$. In terms of irreducible representations of $SO(9)$, there are 128 bosonic degrees of freedom comprised of $44$ graviton degrees of freedom $G_{MN}$ (a symmetric traceless second rank tensor) and $84$ three-form degrees of freedom $B_{MNP}$ (completely antisymmetric rank three tensor). The 128 fermionic degrees of freedom lie in a single Rarita-Schwinger field $\psi_M^\alpha$ ($M$ being the $SO(9)$ index and $\alpha$ the corresponding spinor index). 
\vskip 0.3cm

\ndt We explain briefly how the degrees of freedom in the five-dimensional theory can be understood in terms of those in the eleven-dimensional theory. We split the little group up as $SO(9) \supset SO(3) \times SO(6)$ with $SO(3)$ being the little group in five-dimensions and $SO(6)$, an internal symmetry group. Under this division, we have 
\bea
\bf{44}~=~\bf{21}_{\text{0}} +\bf{18}_{\text{1}}+\bf{5}_{\text{2}}\ ,
\eea
\ndt and
\bea
\bf{84}~=~\bf{20}_{\text{0}}+\bf{1}_{\text{0}}+\bf{45}_{\text{1}}+\bf{18}_{\text{1}},
\eea
\ndt where the subscript denotes the associated $SO(3)$ representation. The internal $SO(6)$ symmetry may be upgraded to an $USp(8)$ symmetry~\cite{Cremmer:1979uq}. This $USp(8)$ group is the maximal compact subgroup of the $E_{6(6)}$ exceptional group which is then the key global symmetry associated with maximal supergravity in five-dimensions. In this paper, we construct the generators for both $USp(8)$ and $E_{6(6)}$ in light-cone superspace and show explicitly the invariance of the supergravity action, to first order in the coupling constant, under these symmetries.
\vskip 0.3cm
\ndt The exceptional symmetries $E_7$ and $E_{8}$ have been extensively studied~\cite{Brink:2008qc,Brink:2008hv,Ananth:2016abv,Ananth:2017nbi} in light-cone superspace and one aim of this paper is to extend this program of study to $E_6$ as well.\\
 
\ndt Rather than attempt a direct formulation of supergravity in five-dimensions or a dimensional reduction of the $d=11$ theory, we start at the well studied four-dimensional theory and use dimensional `oxidation'\cite{Ananth:2005vg} to arrive at the action in $d=5$. The reason we do this is because the $d=4$ formalism is easier to `lift' to $d=5$ and - as we will show - some of the exceptional structures encountered in $d=4$ survive when moving to $d=5$.

\vskip 0.3cm
\ndt We therefore begin with a brief review of the light-cone superspace formulation of four-dimensional $\mathcal{N}=8$ supergravity. 

\section{ (${\mathcal N}=8,d=4)$ supergravity in light-cone superspace}

\ndt Light-cone superspace Grassmann coordinates $\theta^m$ and $\bar{\theta}_m$ which transform as the $8$ and ${\bar 8}$ of SU(8) where ($m=1,2,...,8$) to the bosonic light-cone coordinates 
\bea
{x^{\pm}}=\frac{1}{\sqrt 2}\,(\,{x^0}\,{\pm}\,{x^3}\,)\;;\;\; \quad x =\frac{1}{\sqrt 2}\,(\,x^1\,+\,i\,x^2\,)\;;\;\; \quad {\bar x}=(x)^*\ .
\eea
\ndt All physical degrees of freedom of the ${\mathcal N}=8$ theory are captured by a single complex superfield~\cite{Bengtsson:1983pg}
\bea
\label{superfield}
\begin{split}
\phi\,(\,y\,)\,=&\,\frac{1}{{\parp}^2}\,h\,(y)\,+\,i\,\theta^m\,\frac{1}{{\parp}^2}\,{\bar \psi}_m\,(y)\,+\,\frac{i}{2}\,\theta^m\,\theta^n\,\frac{1}{\parp}\,{\bar A}_{mn}\,(y)\ , \\
\;&-\,\frac{1}{3!}\,\theta^m\,\theta^n\,\theta^p\,\frac{1}{\parp}\,{\bar \chi}_{mnp}\,(y)\,-\,\frac{1}{4!}\,\theta^m\,\theta^n\,\theta^p\,\theta^q\,{\bar D}_{mnpq}\,(y)\ , \\
\;&+\,\frac{i}{5!}\,\theta^m\,\theta^n\,\theta^p\,\theta^q\,\theta^r\,\epsilon_{mnpqrstu}\,\chi^{stu}\,(y)\ ,\\
\;&+\,\frac{i}{6!}\,\theta^m\,\theta^n\,\theta^p\,\theta^q\,\theta^r\,\theta^s\,\epsilon_{mnpqrstu}\,\parp\,A^{tu}\,(y)\ ,\\
\,&+\,\frac{1}{7!}\,\theta^m\,\theta^n\,\theta^p\,\theta^q\,\theta^r\,\theta^s\,\theta^t\,\epsilon_{mnpqrstu}\,\parp\,\psi^u\,(y)\ ,\\
\,&+\,\frac{4}{8!}\,\theta^m\,\theta^n\,\theta^p\,\theta^q\,\theta^r\,\theta^s\,\theta^t\,\theta^u\,\epsilon_{mnpqrstu}\,{\parp}^2\,{\bar h}\,(y)\ .
\end{split}
\eea
\ndt Superspace covariant derivatives $d^m$ and $\bar{d}_{m}$ are defined as
\bea
d^m=-\frac{\del}{\del\bar{\theta}_m}-\frac{i}{\sqrt{2}}\theta^m\del^+\;\;;\;\;\;\;\;\;\bar{d}_m=\frac{\del}{\del\theta^m}+\frac{i}{\sqrt{2}}\bar{\theta}_m\del^+.
\eea
 \ndt The superfield $\phi$ and its complex conjugate $\bar\phi$ satisfy chiral constraints
\be
{d^{m}}\,\phi\,=\,0\ ;\qquad {\bar d_{m}}\,\bar\phi\,=\,0\ ,
\ee
and are related by  ``inside-out" constraints
\be
\bar d_{m}^{}\,\bar d_{n}^{}\,\bar d_{p}^{}\,\bar d_{q}^{}\,\phi~=~\frac{1}{ 2}\,\epsilon_{{m}{n}{p}{q}{r}{s}{t}{u}}^{}\,d^{r}_{}\,d^{s}_{}\,d^{t}_{}\,d^{u}_{}\,\bar\phi\ .
\ee
\vskip 0.3cm
\noindent To order $\kappa$ (the gravitational coupling constant), the ${\mathcal N}=8$ supergravity action reads

\be
\label{cubans}
\int d^4x\int d^8\theta\,d^8 \bar \theta\,{\cal L}\;\equiv\;\int\,{\cal L}\;,
\ee
where
\bea
{\cal L}&=&-\bar\phi\,\frac{\Box}{\partial^{+4}}\,\phi-\;2\,{\kappa}\;{\int}\;{\frac {1}{{\parp}^2}}\;{\nbar \phi}\;\;{\bar \partial}\,{\phi}\;{\bar \partial}\,{\phi}+c.c. ,\nn
\eea

\ndt with Grassmann integration normalized such that $\int d^8\theta\,{(\theta)}^8=1$.\\
\vskip 0.3cm

\ndt On the light-cone, the SuperPoincar\'e algebra splits up into kinematical and dynamical generators. The kinematical generators include
\ndt the three momenta
\be
p^+_{}~=~-i\,\partial^+_{}\ ;\qquad p~=~-i\,\partial\ ; \qquad \bar p~=~-i\,\bar\partial\ ,
\ee
\newline
\ndt the transverse space rotations
\be
j~=~x\,\bar\partial-\bar x\,\partial+ S^{12}_{}\ ,
\ee
where
\be
S^{12}_{}~=~ \frac{1}{4i\sqrt{2}\parp}(q^j\bar{q}_j-\bar{q}_jq^j).
\ee
\ndt and the ``plus-rotations"
\be
j^+_{}~=~i\, x\,\partial^+_{}\ ;\qquad   \bar j^+_{}~=~i\,\bar x\,\partial^+_{}\ ,
\ee
\be
 j^{+-}_{}~=~i\,x^-_{}\,\partial^+_{}-\frac{i}{2}\,(\,\theta^m_{}\frac{\del}{\del\bar{\theta}_m}+\bar\theta^{}_m\,\frac{\del}{\del\theta^m}\,)\ .
\ee
while the dynamical generators are $p^-$, the Hamiltonian and $j^-$, $\bar{j}^-$ the `boosts'. A detailed description of the algebra is given in section 4 of ref.~\cite{Ananth:2005zd}.
\vskip 0.3cm
\ndt The kinematical supersymmetries are given by
\be
\label{su8ks}
q^{\,m}_{\,+}=-\frac{\del}{\del\bar{\theta}_m}\,+\,\frac{i}{\sqrt 2}\,{\theta^m}\,{\partial^+}\ ;\qquad{{\bar q}_{\,+\,m}}=\;\;\;\frac{\del}{\del\theta^m}\,-\,\frac{i}{\sqrt 2}\,{{\bar \theta}_m}\,{\partial^+}\ ,
\ee
\\
\ndt which satisfy
\be
\label{algebra1}
\{\,q^{\,m}_{\,+}\,,\,{{\bar q}_{\,+\,n}}\,\}\,=\,i\,{\sqrt 2}\,{{\delta^m}_n}\,{\parp}\ .
\ee
\ndt These anti-commute with the chiral derivatives 
\bea
\{\,q^{\,m}_{\,+}\,,\,{{\bar d}_n}\,\}\,=\,\{\,{d^m}\,,\,{{\bar q}_{\,+\,n}}\,\}\,=\,0\ ,
\eea
while the commutator of the dynamical supersymmetries $q^{\,m}_{\,-}$ and $\bar{q}_{-\,m}$ close on $p^-$.

\subsection{Exceptional symmetry in $d=4$}

\ndt The four-dimensional theory (\ref{cubans}) admits a global $E_{7(7)}$ symmetry, the generators of which may be written in two parts: those corresponding to its maximal compact subgroup $SU(8)$
\bea
\label{su8}
\delta_{SU(8)}\phi~=~\omega^j_{\;i}\,\frac{1}{i\sqrt{2}\,\del^+}(q^i\,\bar{q}_j-\frac{1}{8}\,\delta^i_j\,q^k \bar{q}_k)\phi\,,
\eea
\ndt where $\omega^j_{\;i}$ ($i,j=1,\ldots,8$) are the 63 parameters of $SU(8)$ and those corresponding to the non-linear coset ${E_{7(7)}/SU(8)}$
\bea
\label{e7}
&&\delta_{E_{7(7)}/SU(8)}\phi=-\frac{2}{\kappa}\theta^{klmn}\overline{\Xi}_{klmn} \\
&&+\;\frac{\kappa}{4!}\Xi^{klmn}\frac{1}{{\del}^{+\,2}}\biggl (\bar{d}_{klmn}\frac{1}{\parp}\phi\,\del^{+\,3}\phi-4\,\bar{d}_{klm}\,\phi\,\bar{d}_n\,\del^{+\,2
}\phi+3\,\bar{d}_{kl}\,\parp\phi\,\bar{d}_{mn}\,\parp\phi\biggr)\ ,\nn
\eea

\ndt where $\Xi^{klmn}$ (the rank four antisymmetric tensor of $SU(8)$) are 70 parameters and satisfy $\Xi^{klmn}=\frac{1}{2}\epsilon^{klmnpqrs}\,\overline{\Xi}_{pqrs}$. 
\vskip 0.1cm
\ndt Invariance of the four-dimensional theory under both the superPoincar\'e symmetry and this exceptional symmetry was shown, in this language, in~\cite{Ananth:2016abv}.
\vskip 0.5cm

\section{Moving up a dimension}

\ndt A key point in this paper is that we will use the {\bf {same}} $(\mathcal N=8, d=4)$ superfield from (\ref {superfield}) to describe the five-dimensional supergravity theory. This is because the five-dimensional theory simply re-packages the degrees of freedom already present in the four-dimensional theory. Specifically, the little group when moving up a dimension is upgraded from $SO(2)$ to $SO(3)$ while the role of $SU(8)$ is played by $USp(8)$, ie. $SU(8)$ representations in the superfield need to be `viewed' as representations of $USp(8)$. Specifically, the $\bf{70}$ of $SU(8)$ splits as
\bea
\bf{70}=\bf{42}+\bf{27}+\bf{1}.
\eea
\ndt We therefore have, in five dimensions, $\bf{42}$ scalar fields, $\bf{27}$ vector bosons and $\bf{1}$ graviton degree of freedom (each with Cartan eigenvalue zero). The $\bf{28}$ of $SU(8)$ splits into 
\bea
\bf{28}~=~\bf{27}+\bf{1}\ ,
\eea
\ndt producing $\bf{27}$ vector bosons and $\bf{1}$ graviton each with Cartan eigenvalue $+1$ (and $-1$ for the $\overline{\bf{28}}$ of $SU(8)$). The $\bf{56}$ of $SU(8)$ is now
\bea
\bf{56}~=~\bf{48}+\bf{8}\ ,
\eea
\ndt representing $\bf{48}$ $\text{spin-}1/2$ and $\bf{8}$ spin-$3/2$ fields with the Cartan eigenvalue $+1/2$ (and $-1/2$ for the $\overline{\bf{56}}$). The physical degrees of freedom in four and five dimensions are summarized below.
\vskip 0.2cm

\begin{longtable} [c] {| c | c | c | c | c | c | c |}
\hline
	Dimension & R-symmetry & Gravitons & Gravitinos & Vectors & Gauginos & Scalars\\
\hline
	4   &  SU(8) &  2 & 16 & 56 & 112 & 70\\
\hline
	5  &  USp(8) &  5 & 32 & 81 & 96 & 42\\
\hline
\end{longtable}

\vskip 0.3cm

\subsection{Grassmann variables and spinors in five dimensions}

\ndt Theories of supergravity with extended supersymmetry in five-dimensions have the symplectic group $USp(2n)$ as their $R$-symmetry group~\cite{Cremmer:1980gs} (a light-cone related reference on this group is~\cite{Brink:2010ti}). The $R$-symmetry group relevant to the present paper is $USp(8)$ with a $36$-dimensional Lie algebra. The little group in five-dimensions is $SO(3)\sim SU(2)$. We thus introduce the symplectic Grassmann variables $\theta^{i\alpha}$ (where $i=1,2,..8$ is now an $USp(8)$ index and $\alpha =1,2$ is the $SU(2)$ index) satisfying
\bea
\label{symplectic}
\bar{\theta}_{i\alpha}~=~\theta^{j\beta}\,C_{ji}\,\epsilon_{\beta\alpha}\, ,
\eea 

\ndt where $\epsilon_{\alpha\beta}$ (antisymmetric in $\alpha\,,\beta$) is the invariant tensor of $SU(2)$. The $C_{ij}$ is the symplectic invariant (or the charge-conjugation matrix), antisymmetric and satisfying $C^{ij}C_{jk}=-\delta^{i}_{\,k}$.\\

\ndt The algebra satisfied by the Grassmann variables is 
\bea
\{\theta^{i\alpha},\theta^{j\beta}\}~=~0\;\;\;\;;\;\;\;\;\{\theta^{i\alpha},\frac{\del}{\del \theta^{j\beta}}\}~=~\delta^i_{\,j}\,\delta^{\alpha}_{\,\beta}\ .
\eea

\ndt One can define kinematical supersymmetries
\bea
\label{susy}
q^{i\alpha}~&=&~-\frac{\del}{\del \bar{\theta}_{i\alpha}}+\frac{i}{\sqrt{2}}\theta^{i\alpha}\del^+\,,\nn\\
\bar{q}_{j\beta}~&=&~\frac{\del}{\del \theta^{j\beta}}-\frac{i}{\sqrt{2}}\bar{\theta}_{j\beta}\del^+\ ,
\eea
\ndt which satisfy the algebra 
\bea
\label{algebra2}
\{q^{i\alpha},\bar{q}_{j\beta}\}~=~i\sqrt{2}\delta^i_{\,j}\,\delta^{\alpha}_{\,\beta}\;\;\;\;;\;\;\;\;\{q^{i\alpha},q^{j\beta}\}~=~-i\sqrt{2}\,C^{ij}\,\epsilon^{\alpha\beta}.
\eea
\ndt Since the second component of $\theta ^{j\alpha}$ ($\alpha=2$) is related to the first component of $\bar{\theta}_{i\beta}$ ($\beta=1$) ($\bar{\theta}_{i1}=C_{ij}\theta^{j2}$ from (\ref{symplectic})), we can simply work with the first component ($\theta^{i1}\equiv\theta^i$)~\cite{Brink:2010ti}.  It is then easy to verify that the $q^{i1}\,,\bar{q}_{j1}$ are the $SU(8)$ kinematical supersymmetries defined in equation (\ref{su8ks}) of section 3. This is due to the fact that the algebra (\ref{algebra2}) with $\alpha,\beta=1$, reduces to that of the $SU(8)$ supercharges (\ref{algebra1}). Hence, it is possible to continue to work with the $SU(8)$ supercharges in five-dimensions.

\vskip 0.3cm

\subsection{Five dimensions}

\vskip 0.3cm

\ndt The goal in this section is to begin with (\ref {cubans}) and `lift' it up one dimension so we have a five-dimensional theory with the same degrees of freedom - but re-packaged appropriately given the symmetries of the five-dimensional spacetime.
\vskip 0.1cm
\ndt To move up one dimension, we introduce an additional coordinate $x^5$ and its derivative $\del_5$. The little group in five dimensions is $SO(3)$ so we simply add to the four-dimensional little group $SO(2)$, the coset generators of $SO(3)/SO(2)$~\cite{Brink:2010ti}
\bea
j^5~=~i(x^5\del-x\del^5)-\frac{1}{4i\del^+}q^k\,C_{kl}q^l\ ,\nn\\
\bar{j}^5~=~i(x^5\bar{\del}-\bar{x}\del^5)+\frac{1}{4i\del^+}\bar{q}_k\,C^{kl}\bar{q}_l.
\eea
\ndt The spin part of the generators is fixed using three conditions. The consistency with helicity and the  requirement that the spin variation of the superfield commute with the chiral derivatives fixes the spin part to be a quadratic function of the kinematical supersymmetries ($q$'s in the case of $j^5$ and $\bar{q}$'s in the case of $\bar{j}^5$). Finally, the dimension of the generator fixes the $\del^+$ structure. 
\vskip 0.3cm
\ndt The $SO(3)$ algebra is spanned by $j\,,\,j^{5}\,,\,\bar{j}^{5}$ which obey
\bea
[j^5,\bar{j}^5]~=~j\;;\;\;\;\;\;\;\;\;[j,j^5]~=+j^5\;;\;\;\;\;\;\;\;[j,\bar{j}^5]~=-\bar{j}^5.
\eea

\ndt As a first step, we allow the superfield to depend on the extra coordinate $x^5$. As shown in~\cite{Ananth:2005vg}, covariance may be established by replacing the $SO(2)$ derivatives by the generalized derivative $\overline{\nabla}$ and its conjugate. For five dimensions, we conjecture the following form for the generalized derivative 
\bea
\overline{\nabla}=\bar{\del}+\frac{\sigma}{16}\,\bar{d}_m\,C^{mn}\,\bar{d}_n\frac{\del^5}{\parp}.
\eea
\ndt We verify that this conjecture is okay by performing two rotations as shown below (with the coset transformations $j^5$ and $\bar{j}^5$)
\bea
[\,\overline{\nabla},j^5]=\nabla^5=-i\del^5+i\frac{\sigma}{16}\bar{d}_m\,C^{mn}\,\bar{d}_{n}\frac{\del}{\parp}\ ,
\eea
and
\bea
[\nabla^5,\bar{j}^5]=\overline{\nabla}\ ,
\eea
\ndt assuring us that $(\overline{\nabla},\nabla^5)$ transforms appropriately in five dimensions. The exact value of the constant $\sigma$ will now be determined by demanding Lorentz invariance of the cubic vertex (obtained by replacing $\bar{\del}$ by $\overline{\nabla}$) in $d=5$, the focus of the next sub-section.

\vskip 0.3cm

\subsection{Invariance of the action}

The kinetic term, in (\ref{cubans}) is rendered $SO(3)$-invariant by simply including the extra transverse derivative in the d'Alembertian. Moving to the cubic vertex in (\ref {cubans}), the proposal is that it remains the same for the five-dimensional theory except for the replacement $\partial, \bar\partial\;\rightarrow\;\nabla,\overline{\nabla}$ (and the added dependence of the superfields on the fifth coordinate), ie.
\bea
\label{fived}
\begin{split}
{\mathcal V}\;=&-\;2\,{\kappa}\;{\int}\,{d^{5}}x\,{\int}\,{d^8}{\theta}\,{d^8}{\bar \theta}\;{\frac {1}{{\parp}^2}}\;{\nbar \phi}\;\;{\nbar \nabla}\,{\phi}\;{\nbar \nabla}\,{\phi}\ ,\\ 
\end{split}
\eea
together with its complex conjugate. To show $SO(3)$ invariance, we consider the variations
\bea
{\delta_{j^5}}\,{\phi}\;=\;\,\frac{i}{4\parp}\;{\bar{\omega}_5}\,\,\;{q^m}\,\,{C}_{mn}\,{q^n}\,{\phi}\ ,
\eea

\bea
{\delta_{j^5}}\,{\bar{\phi}}\;=\;\,\frac{i}{4\parp}\;{\bar{\omega}_5}\,\,\;{q^m}\,\,{C}_{mn}\,{q^n}\,{\bar{\phi}}\ ,
\eea

\bea
{\delta_{j^5}}\,{\nbar \nabla}\;=\;-\,{\bar{\omega}_5}\,{\nabla^5}\ ,
\eea
where $\bar{\omega}_5$ is the parameter corresponding to the coset transformation $j^5$.  We show the invariance under the coset $SO(3)/SO(2)$ transformations (since invariance under the $SO(2)$ is clear), which proves the invariance under the full $SO(3)$.
\vskip 0.3cm
\ndt The contributions from the variation of the superfields and the generalised derivative are given below (terms that involve two $SO(2)$ derivatives or two $\del^5$s, cancel trivially). The non-trivial terms all involve a single $SO(2)$ derivative and a single ${\partial^5}$.
\vskip 0.1cm
\ndt {\bf {from}} $\int\;{\frac {1}{{\parp}^2}}\;{\biggl (}\,{\delta_{j^5}}\,{\nbar \phi}\,{\biggr )}\;{\nbar \nabla}\,{\phi}\;{\nbar \nabla}\,{\phi}$:
\bea
\begin{split}
\int\;\frac{i\,\sigma}{4}{\biggl \{}&{\frac {8}{{\parp}^3}}\,{\nbar \phi}\;{\bar \partial}\,{\phi}\,{\partial^5}\,{\parp}\,{\phi}\;-\;{\frac {4}{{\parp}^3}}\,{\nbar \phi}\;{\bar \partial}\,{\phi}\,{\partial^5}\,{\parp}\,{\phi} ,\\
&+\;{\frac {1}{{\parp}^3}}\,{\nbar \phi}\;{\bar \partial}\,\phi\,{\theta^m}\,{{\bar d}_m}\,{\partial^5}\,{\parp}\,{\phi}, \\
&+\,{\frac {1}{{\parp}^3}}\,{\nbar \phi}\;{\theta^m}\,{\bar \partial}\,{\parp}\,{\phi}\,\,{{\bar d}_m}\,{\partial^5}\,{\phi}\;\biggr \}.
\end{split}
\eea
\vskip 0.3cm

\ndt {\bf {from}} $\int\;{\frac {1}{{\parp}^2}}\;{\nbar \phi}\,\;{\biggl (}\,{\delta_{j^5}}\,{\nbar \nabla}\,{\biggr )}\,{\phi}\;{\nbar \nabla}\,{\phi}$:

\bea
\int\;{\frac {2\,i\,}{{\parp}^2}}\,{\nbar \phi}\;{\bar \partial}\,{\phi}\,{\partial^5}\,{\phi}.
\eea

\vskip 0.3cm

\ndt {\bf {from}} $\int\;{\frac {1}{{\parp}^2}}\;{\nbar \phi}\,\;{\nbar \nabla}\,{\biggl (}\,{\delta_{j^5}}\,{\phi}\,{\biggr )}\;{\nbar \nabla}\,{\phi}$:

\bea
\int\;\frac{i\,\sigma}{4}\,{\biggl \{}\;{\frac {4}{{\parp}^2}}\,{\nbar \phi}\;{\bar \partial}\,{\phi}\,{\partial^5}\,{\phi}\;+\;{\frac {1}{{\parp}^2}}\,{\nbar \phi}\;{\theta^m}\,{\bar \partial}\,{\phi}\,\,{{\bar d}_m}\,{\partial^5}\,{\phi}\;{\biggr \}}.
\eea

\vskip 0.3cm
\noindent A useful identity when simplifying this is 
\bea
{\int}\,{\frac {1}{{\parp}^3}}\,{\nbar \phi}\,{\bar \partial}\,{\phi}\,{\partial^5}\,{\parp}\,{\phi}\,=\,{\int}\,{\frac {1}{{\parp}^3}}\,{\nbar \phi}\,{\partial^5}\,{\phi}\,{\bar \partial}\,{\parp}\,{\phi}\ ,
\eea
which is a consequence of the inside-out constraints (plus partial integrations). We find
\bea
{\delta_{j^5}}\,{\cal V}\,\propto\,\int\,{\biggl (}\;\frac{i\,\sigma}{2}+\,2i\,{\biggr )}\;{\frac {1}{{\parp}^2}}\,{\nbar \phi}\;{\bar \partial}\,{\phi}\,{\partial^5}\,{\phi}\ .
\eea
\noindent As stated already, the invariance requires this variation to vanish so that $\sigma=-4$ and
\bea
{\nbar \nabla}\;=\;{\bar \partial}\,-\,{\frac {1}{4}}\,{{\bar d}_m}\,{{C}^{mn}}\,{{\bar d}_n}\,{\frac {\partial^5}{\parp}}\ ,
\eea
completing the proof of $SO(3)$ invariance of our ansatz for the cubic vertex in (\ref {fived}). 
\noindent In this light-cone form, the Lorentz invariance in five-dimensions is automatic, once the invariance under the little group has been proven.

\section{$E_{6(6)}$}

\ndt In five-dimensions, the relevant global symmetry group is $E_{6(6)}$ which has $78$ generators. The Lie algebra splits into that of the $USp(8)$ and the coset $E_{6(6)}/USp(8)$ so we have
\bea
\bf{78}~=~\bf{36}+\bf{42}.
\eea

\ndt The $\bf{36}$ linear (symmetric) generators are constructed from the supercharges $q$ and $\bar{q}$
\bea
T^{ij}=\frac{1}{i\sqrt{2}\parp}(q^i\,C^{jk}\,\bar{q}_k\,+\,q^j\,C^{ik}\,\bar{q}_k)\ ,
\eea
and satisfy the algebra
\bea
[T^{ij},T^{mn}]~=~C^{jm}\,T^{in}+C^{jn}\,T^{im}+C^{im}\,T^{jn}+C^{in}\,T^{jm}\ .
\eea

\ndt On the superfield the action of the $USp(8)$ generators is 

\bea
\delta_{USp(8)}\phi~=~\omega_{ij}\,T^{ij}\phi\, ,
\eea

\ndt where $\omega_{ij}$ are the $\bf{36}$ parameters.\\

\ndt  The $\bf{42}$ coset transformations act non-linearly on the scalars of the theory~\cite{Cremmer:1979up}. These are of the form $a\kappa^{-1}+b\kappa+\mathcal{O}(\kappa^3)$. The $\bf{42}$ of $USp(8)$ is given by a completely anti-symmetric traceless rank four tensor. We thus conjecture that the lowest order ($\kappa^{-1}$) transformations are

\bea
\label{nl}
\delta_{E_{6(6)}/USp(8)}\phi&=&-\frac{2}{\kappa}\theta^{klmn}\overline{\Sigma}_{klmn},
\eea 

\ndt where $\theta^{klmn}=\theta^k\theta^l\theta^m\theta^n$ and $\overline{\Sigma}_{klmn}$ are the $\bf{42}$ parameters which like the scalar fields satisfy ${\Sigma}^{klmn}=\Sigma_{klmn}^*\equiv\overline{\Sigma}_{klmn}$.  At this order, it is only the scalar fields which are affected by these transformations- they act like a shift symmetry. This variation (\ref{nl}) is fixed due to the constraints imposed by helicity and dimension.\\

\ndt The generators at order $\kappa$ are fixed by the requirement that the commutator of two coset generators close on an $USp(8)$ generator. We find
\bea
\label{e6}
&&\delta_{E_{6(6)}/USp(8)}\phi=-\frac{2}{\kappa}\theta^{klmn}\overline{\Sigma}_{klmn} \\
&&+\;\frac{\kappa}{4!}\Sigma^{klmn}\frac{1}{{\del}^{+\,2}}\biggl (\bar{d}_{klmn}\frac{1}{\parp}\phi\,\del^{+\,3}\phi-4\,\bar{d}_{klm}\,\phi\,\bar{d}_n\,\del^{+\,2
}\phi+3\,\bar{d}_{kl}\,\parp\phi\,\bar{d}_{mn}\,\parp\phi\biggr)\ ,\nn
\eea

\ndt where $\bar{d}_{i_1i_2..i_n}=\bar{d}_{i_1}\bar{d}_{i_2}....\bar{d}_{i_n}$. It can be checked that the above variation is chiral.
\vskip 0.3cm
\ndt With $\boldsymbol{\delta}\phi ~\equiv~\delta_{E_{6(6)}/USp(8)}$ and $\delta\phi~\equiv~\delta_{USp(8)}\phi$ we have, schematically, for the $E_{6(6)}$ algebra
\bea
[\boldsymbol{\delta}_1,\boldsymbol{\delta}_2]\phi~=~\delta\phi\;\;\;\;\;\;;\;\;\;\;\;\;[\boldsymbol{\delta},\delta]\phi~=~\boldsymbol{\delta}\phi\;\;\;\;\;\;;\;\;\;\;\;\;[\delta_1,\delta_2]\phi~=~\delta\phi\ ,
\eea
closely resembling the $E_{7(7)}$ results in four dimensions~\cite{Brink:2008qc}.
\vskip 0.3cm
\ndt The five-dimensional action to the cubic order is trivially invariant under the linear $USp(8)$ generators. Since the order 
$\frac{1}{\kappa}$ variation maps a superfield to a constant, it can be shown that the action is invariant to this order. 
\vskip 0.3cm

\ndt{\bf{Quartic vertex}}
\vskip 0.3cm
\ndt To verify invariance of the action at the next order, ie. order $\kappa$, we note that there are two contributions: the order $\kappa$ variation of the kinetic term and the order $\frac{1}{\kappa}$ variation of the quartic vertex. However, the complete quartic interaction vertex in light-cone superspace is not known - for the $\mathcal N=8$ theory. 
\vskip 0.3cm
\ndt There have been several attempts made to derive the quartic interaction vertices in light-cone superspace~\cite{Brink:2008qc,Brink:2008hv,Ananth:2006fh}. However, the expressions derived are very cumbersome (hundreds of terms even in superspace) making them unsuitable for use in explicit calculations. However, we explain in the following section how a compact expression for these quartic vertices may be arrived at.

\section{Leveraging exceptional symmetries}

\ndt In the light-cone gauge, Lorentz invariance is no longer manifest. Demanding closure of the Lorentz commutators, then serves as a tool to derive interaction vertices. This approach is useful but can encounter challenges when applied to quartic and higher-point vertices. We describe in  this section, an alternate way of deriving interaction vertices making use of the exceptional symmetry~\cite{Ananth:2016abv}.
\vskip 0.3cm
\ndt Since the non-linear exceptional symmetry appears as a series in $\kappa$ with only the odd powers, it relates higher {\it odd} ({\it even}) point vertices to lower {\it odd} ({\it even}) point vertices as shown in the figure. Hence, if we use these to determine interaction vertices, then the role of the dynamical generators in the usual light-cone procedure is played instead by the exceptional symmetries. Schematically we have
\vskip 0.3cm
\begin{center}
\begin{figure}[h]
\begin{center}
\begin{tikzpicture}[node distance=2cm]
\node (k) [forces] {kinetic};
\node (c) [forces, right of=k, xshift=3cm] {cubic};
\node (qa) [forces, above of=k, yshift=1cm] {quartic};
\node (qi) [forces, right of=qa, xshift=3cm] {quintic};
\node (6)  [forces, above of=qa, yshift=1cm ] {6-point};
\node (7) [forces, right of=6, xshift=3cm] {7-point};
\draw [arrow1] (6) -- (7);
\draw [arrow1] (qi) -- (6);
\draw [arrow] (qa) -- (6);
\draw [arrow] (qi) -- (7);
\draw [arrow1] (k) --  (c);
\draw [arrow1] (qa) --  (qi);
\draw [arrow1] (c) -- (qa);
\draw [arrow] (k) -- (qa);
\draw [arrow] (c) -- (qi);
\end{tikzpicture}
\end{center}
\caption{The red-arrows indicate the usual light-cone approach to deriving interaction vertices using dynamical generators while the black arrows indicate the alternate path, using the non-linearly realized exceptional symmetries}
\end{figure}
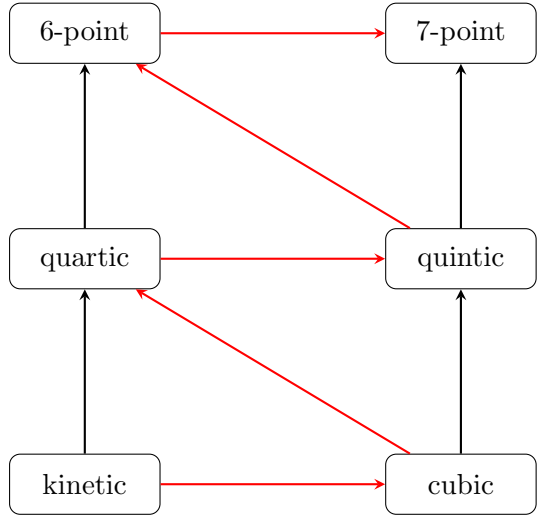
\end{center}
\vskip 0.3cm
\ndt Note that in the case of maximal supergravity in three-dimensions, only even-point vertices exist and there are no odd-point vertices. Thus, the two approaches coincide in that case.
\vskip 0.3cm
\ndt Requiring invariance of the supergravity Hamiltonian under the exceptional symmetry group in $d$ dimensions, $3\le d \le5$, yields the following bootstrap equation
\bea
\label{bootstrap}
\delta_{1/{\kappa}}(\mathcal{V}_4)~=~-\delta_{\kappa}(\mathcal{V}_2),
\eea

\ndt where $\mathcal{V}_4$ is the unknown quartic vertex while $\mathcal{V}_2$ is the kinetic term in the Hamiltonian. $\mathcal{V}_4$ may be written as the sum of a homogeneous piece and an inhomogeneous piece
\bea
\mathcal{V}_4=\mathcal{V}_4'+\mathcal{V}_4''\ ,
\eea
\ndt such that 
\bea
\delta_{1/\kappa}\mathcal{V}_4'=0\ .
\eea
This $\mathcal{V}_4'$ piece is easy to write down because any quartic vertex with a spacetime derivative on each superfield will satisfy this relation (since the order $\fr{\kappa}$ maps a superfield to a constant). We also have two additional conditions
\bea
\label{inh}
\delta_{1/{\kappa}}(\mathcal{V}_4'')=-\delta_{\kappa}(\mathcal{V}_2)\ ,
\eea
and 
\bea
\label{boundary}
\mathcal{V}_4|_{\text{\{all fields except graviton set to zero\}}}~=~\mathcal{M}_4\ ,
\eea

\ndt where $\mathcal{M}_4$ is the quartic interaction vertex for pure gravity in the light-cone gauge~\cite{Bengtsson:1983vn,Ananth:2006fh}. 

\subsection{Four dimensions}
\vskip 0.3cm

\ndt In four-dimensions, the first step is to find the inhomogeneous quartic pieces $\mathcal{V}_4''$ using $E_{7(7)}$. We can then make use of the kinematical symmetry generators to find the homogeneous pieces. The Hamiltonian of the $d=4$ theory is
\bea
\label{H}
\mathcal{H}~=~\int\,d^3 x\, d^8\theta\, d^8\bar{\theta}\left[\;2\;\bar{\phi}\;\frac{\del\bar{\del}}{\del^{+\,4}}\;\phi\;+\;2\,{\kappa}\;\left({\frac {1}{{\parp}^2}}\;{\nbar \phi}\;\;{\bar \partial}\,{\phi}\;{\bar \partial}\,{\phi}+c.c.\right)\right]\,.
\eea

\ndt We use the generators of $E_{7(7)}$  presented in equation (\ref{e7}) of section $3$.
Varying $\phi$ in the kinetic term of (\ref{H}) with respect to the non-linear $E_{7(7)}$ symmetry yields a contribution at order $\kappa$ (since the order $\frac{1}{\kappa}$ variation vanishes)
\bea
2\,\frac{\kappa}{4!}\Xi^{klmn}\,\frac{\del\bar{\del}}{\del^{+\,6}}\,\bar{\phi}\,\biggl (\frac{1}{\parp}\phi\,\bar{d}_{klmn}\del^{+\,3}\phi+4\,\,\phi\,\bar{d}_{klmn}\,\del^{+\,2
}\phi+3\,\,\parp\phi\,\bar{d}_{klmn}\,\parp\phi\biggr)\,,
\eea
\ndt which is obtained after performing some partial integrations (the measures and integrals are suppressed). This further simplifies to
\bea
\label{junk}
&&2\,\frac{\kappa}{4!}\Xi^{klmn}\,\biggl(2\,\frac{\del\bar{\del}}{\del^{+\,4}}\,\bar{\phi}\,\frac{1}{\del^+}\phi\,\bar{d}_{klmn}\parp\phi\\
&&\hspace{3.5cm}\,-\,\,\frac{\del\bar{\del}}{\del^{+\,6}}\,\bar{\phi}\,\frac{1}{\parp}\phi\,\bar{d}_{klmn}\del^{+\,3}\phi\,+\,\frac{\del\bar{\del}}{\del^{+\,6}}\,\bar{\phi}\,\parp\phi\,\bar{d}_{klmn}\,\parp\phi\biggr)\,.\nn\\\nn
\eea
\ndt Varying the superfield $\bar{\phi}$ in the kinetic term of (\ref{H}) yields the complex conjugate of the above contribution. The inhomogeneous part $\mathcal{V}_4''$ is then given by
\bea
\label{v4}
\mathcal{V}_4''~&=&~+\frac{1}{4!}\,\kappa^2\biggl[2\,\frac{\del\bar{\del}}{\del^{+\,4}}\,\bar{\phi}\,\frac{1}{\del^+}\phi\;\bar{d}_{klmn}\parp\phi\\
&&\,-\,\,\frac{\del\bar{\del}}{\del^{+\,6}}\,\bar{\phi}\,\biggl(\,\frac{1}{\parp}\phi\,\bar{d}_{klmn}\del^{+\,3}\phi\,-\,\parp\phi\,\bar{d}_{klmn}\,\parp\phi\biggr)\biggr]\,d^{klmn}\,\bar{\phi}\,+\,c.c.\,. \,\,\nn\\\nn
\eea
\ndt It is straightforward to verify that this result is consistent with the generators $j,j^{+-}$. 
\vskip 0.1cm
\ndt It is equally straightforward to check that this expression (\ref {v4}) is not invariant under the remaining kinematical generators, $j^+,\bar{j}^+$.  To find the homogeneous part of $\mathcal{V}_4$, we ask for invariance of (\ref{v4}) under these two kinematic generators. This produces additional terms and results in a quartic interaction vertex
\bea
\label{v41}
{\mathcal V}_4~&=&~+\frac{1}{4!}\,\kappa^2\biggl[2\,\frac{\del\bar{\del}}{\del^{+\,4}}\,\bar{\phi}\,\frac{1}{\del^+}\phi\;\bar{d}_{klmn}\parp\phi\\
&&\,-\,\,\frac{\del\bar{\del}}{\del^{+\,6}}\,\bar{\phi}\,\biggl(\,\frac{1}{\parp}\phi\,\bar{d}_{klmn}\del^{+\,3}\phi\,-\,\parp\phi\,\bar{d}_{klmn}\,\parp\phi\biggr)\biggr]\,d^{klmn}\,\bar{\phi}\nn\\
&&-\frac{1}{4!}\,\kappa^2\biggl[2\,\frac{\del}{\del^{+\,3}}\,\bar{\phi}\,\frac{1}{\del^+}\phi\;\bar{d}_{klmn}\parp\phi\nn\\
&&-\,\,\frac{\del}{\del^{+\,5}}\,\bar{\phi}\,\biggl(\,\frac{1}{\parp}\phi\,\bar{d}_{klmn}\del^{+\,3}\phi\,-\,\parp\phi\,\bar{d}_{klmn}\,\parp\phi\biggr)\biggr]\,\frac{\bar{\del}}{\parp}d^{klmn}\,\bar{\phi}\nn\\
&&-\frac{1}{4!}\,\kappa^2\biggl[2\,\frac{\bar{\del}}{\del^{+\,3}}\,\bar{\phi}\,\frac{1}{\del^+}\phi\;\bar{d}_{klmn}\parp\phi\nn\\
&&-\,\,\frac{\bar{\del}}{\del^{+\,5}}\,\bar{\phi}\,\biggl(\,\frac{1}{\parp}\phi\,\bar{d}_{klmn}\del^{+\,3}\phi\,-\,\parp\phi\,\bar{d}_{klmn}\,\parp\phi\biggr)\biggr]\,\frac{\del}{\parp}d^{klmn}\,\bar{\phi}\nn\\
&&+\frac{1}{4!}\,\kappa^2\biggl[2\,\frac{1}{\del^{+\,2}}\,\bar{\phi}\,\frac{1}{\del^+}\phi\;\bar{d}_{klmn}\parp\phi\nn\\
&&\,-\,\,\frac{1}{\del^{+\,4}}\,\bar{\phi}\,\biggl(\,\frac{1}{\parp}\phi\,\bar{d}_{klmn}\del^{+\,3}\phi\,-\,\parp\phi\,\bar{d}_{klmn}\,\parp\phi\biggr)\biggr]\,\frac{\del\bar{\del}}{\del^{+\,2}}d^{klmn}\,\bar{\phi}\,+c.c.\ ,\nn
\eea

\ndt which is consistent with the exceptional symmetry and all the kinematical symmetries. It is a sum of homogeneous and inhomogeneous pieces. When one derives a light-cone vertex, one first fixes the form of the vertex using kinematical generators. The dynamical generators then fix the $\del^+$ structure. In this case, the $\del^+$ structure is already fixed as it is inherited from the kinetic term due to the exceptional symmetry  thus suggesting that the vertex (\ref{v41}) is consistent with the dynamical generators. Finally, to the point of whether there could be additional homogeneous terms, which do not mix with the above terms, the boundary condition (\ref{boundary}) would serve as a check.\\

\ndt We can also determine the inhomogeneous part of the quintic vertices~\cite{Ananth:2008ik}, using this procedure starting from the cubic vertices in (\ref{H}). The order $\frac{1}{\kappa}$ variation of the cubic vertices vanishes so the order $\kappa$ variation must be cancelled by the order $\frac{1}{\kappa}$ variation of the inhomogeneous quintic vertex. Accordingly, we find
\bea
\label{v5}
\mathcal{V}_5''&~=~& -2\;\frac{\kappa^3}{ 4!}\,\biggl\{\sum_{n=0}^{4}\,{4\choose n}\;\;\frac{\bar{\del}}{\del^{+\,2}}\left[\frac{1}{\del^{+\,2}}\bar{\phi}\,\bar{\del}\,\bar{d}^{\,n}\phi\right]\frac{1}{\del^+}\phi\,\del^{+\,3}\,\bar{d}^{\,4-n}\phi\\
&&+\;\biggl(2\,\sum_{n=0}^{3}\,{3\choose n}\,\bar{\del}\left[\frac{1}{\del^{+\,2}}\bar{\phi}\,\bar{\del}\,\bar{d}^{\,n}\phi\right]\frac{1}{\del^+}\phi\,\del^{+}\,\bar{d}^{\,4-n}\phi\nn\\
&& -\;2\,\frac{\bar{\del}}{\del^{+\,2}}\left[\frac{1}{\del^{+\,2}}\bar{\phi}\,\bar{\del}\,\bar{d}^{\,n}\phi\right]\,\left[\del^+\phi\,\del^+\,\bar{d}^{\,4-n}\phi\,+\,\frac{1}{\del^+}\phi\,\del^{+\,3}\,\bar{d}^{\,4-n}\phi\right]\biggr)\nn\\
&&+\;3\,\frac{\bar{\del}}{\del^{+\,2}}\left[\frac{1}{\del^{+\,2}}\bar{\phi}\;\bar{\del}\,\phi\right]\,\bar{d}_{ij}\,\del^+\,\phi\,\bar{d}_{kl}\,\del^+\,\phi\biggr\}\,d^{ijkl}\,\bar{\phi}\nn\\
&&+\;\frac{\kappa^3}{ 4!}\biggl[2\,\frac{1}{\del^{+\,2}}(\,\del \,\bar{\phi}\,\del \,\bar{\phi}\,)\,\frac{1}{\del^+}\phi\;\bar{d}_{klmn}\parp\phi\nn\\
&&-\frac{1}{\del^{+\,4}}(\del \,\bar{\phi}\;\del \,\bar{\phi})\biggl(\,\frac{1}{\parp}\phi\,\bar{d}_{klmn}\del^{+\,3}\phi\,-\,\parp\phi\,\bar{d}_{klmn}\,\parp\phi\biggr)\biggr]\,d^{klmn}\,\bar{\phi}\,+\,c.c. \,\,.\nn
\eea

\ndt In principle, one can then use the kinematical generators to determine the homogeneous pieces (deriving six-point and higher vertices~\cite{Ananth:2022spf} would need expressions for the coset transformations to order $\kappa^3$). 
\vskip 0.3cm

\ndt  The same procedure extends nicely to five dimensions with the relevant exceptional symmetry being $E_{6(6)}$. We find that the quartic inhomogeneous piece has the same form as (\ref{v4}) (with $\del\bar{\del}$ replaced by $\del\bar{\del}+\frac{1}{2}\del^5\del^5$). As in four dimensions, the kinematic generators will then provide the homogenous piece. In a similar manner, the five-dimensional quintic vertices may be obtained starting from the corresponding cubic vertices in (\ref{fived}).
\vskip 0.3cm
\ndt Exceptional symmetries play a key role in supergravity theories in many ways and specifically, as discussed here, relate higher-point interaction vertices to lower-point ones. The derivation of the vertices (\ref{v4}), (\ref{v41}) and (\ref{v5}) is a first step in illustrating the power of exceptional symmetries in this regard. A key motivation to study exceptional symmetries remains the issue of ultra-violet (UV) finiteness. Is the $\mathcal N=8$ model actually an example of a UV finite quantum theory of gravity? In~\cite{Bern:2018jmv}, it was shown, up to 4-loops, that the critical dimension for UV divergences to appear in maximal supergravity is
\bea
\label{critical}
d_c~=~\frac{6}{L} + 4\ , 
\eea
\ndt $2\leq L\leq 4$ being the number of loops. It was also shown that this result does not hold at 5-loops, confirming the predictions in~\cite{Vanhove:2010nf,Bjornsson:2010wm}.
\vskip 0.3cm
\ndt Given that light-cone superspace was ideally suited for a proof of UV finiteness of $\mathcal{N}=4$ superYang-Mills theory, it is natural to ask what the formalism can teach us about the UV properties of maximal supergravity theories - armed with the exceptional symmetry machinery presented here. 
\vskip 0.3cm
\ndt The construction of counter-terms in non-supersymmetric theories formulated in the light-cone gauge was discussed in~\cite{Bengtsson:2012dw, Ananth:2023qrf}. An interesting direction to follow up on would be to examine counter-terms in the light-cone superspace theories described in this paper. 

\vskip 0.3cm

\section*{Acknowledgement}
\ndt We thank Sucheta Majumdar and Saurabh Pant for helpful discussions. NB acknowledges a CSIR-NET fellowship.

\end{document}